\documentclass[ajp,twocolumn,letterpaper,noeprint,nodoi,footinbib,superscriptaddress]{revtex4-2} 

\usepackage{amsmath}  
\usepackage{amsfonts} 
\usepackage{graphicx} 
\usepackage{xspace}

\newcommand{\HI}{\ensuremath{\text{H\,I}}\xspace}
\newif\ifanon
\anonfalse 

\begin{document}


\title{Determination of Earth's Orbital Parameters \\ Using Doppler 
Measurements}

\ifanon 
   \author{Anonymous}
   \affiliation{Not specified}
\else 
   \author{Gary Atkins}
   \affiliation{Canadian Centre for Experimental Radio Astronomy, Carp, ON, Canada}
   \author{Marcus Leech}
   \affiliation{Canadian Centre for Experimental Radio Astronomy, Carp, ON, Canada}
   \author{Daniel Marlow} 
   \affiliation{Canadian Centre for Experimental Radio Astronomy, Carp, ON, Canada}
   \affiliation{Physics Department, Princeton University, Princeton NJ, USA} 
   \author{Doug Yuill}
   \affiliation{Canadian Centre for Experimental Radio Astronomy, Carp, ON, Canada}
\fi 

\date{\today}

\begin{abstract} Precision determinations of the parameters of Earth's orbit around the Sun
using methods based on the Doppler shift of 21-cm emissions from the galaxy and the 
repetition rate of pulsar J0332+5434 are reported.  The observed values and statistical errors 
for the orbital parameters for the 21-cm observations are semi-major axis 
$a = 1.00020 \pm 0.00015~{\rm au}$ and eccentricity 
$e= 0.017077 \pm 0.000077$.    For the pulsar observations, the observed values are  
$a = 1.000034 \pm 0.000034~{\rm au}$ and 
$e=0.016667 \pm 0.000017$.  
The observations are based on a year-long campaign 
of daily drift-scan observations using a 12.8-m parabolic antenna and a receiver chain tuned 
to the 21-cm frequency band.    The data used in the analysis are available  for use in 
educational exercises in the supplementary material.\footnote{This article has been submitted to the American Journal of Physics.} 

\end{abstract}

\maketitle

\section{Introduction}
The elliptical orbits that appear as a solution to the Kepler problem are a staple 
of most introductory courses in classical dynamics.  Given the complexity involved in
acquiring observational data related to these orbits, the treatment is generally limited 
to theory only.   In this paper, we outline experimental methods that can be used to establish the 
elliptical nature of Earth's orbit about the Sun, and we use those methods to determine orbital 
parameters, such as the semi-major axis $a$ and the eccentricity $e$.   The time and angle of 
perihelion can also be determined.   

The use of Doppler observations of 21-cm emissions to make a precise determination of the astronomical unit was 
first presented by Knowles\cite{Knowles69}.   In the same paper, Knowles also anticipated the use 
of pulsar timing for the purpose of determining orbital parameters.  The analysis presented here, which 
is designed to be accessible to advanced undergraduates, is a simplified version of more elaborate modern-day 
analyses that use large pulsar observation samples from research radio telescopes to constrain planetary masses 
(see e.g., Ref.~\cite{Champion10}). 

Although celestial orbits are typically discussed in terms of position versus time, $\vec r(t)$, they
can equally well be characterized in terms of velocity versus time, $\vec v(t)={\rm d}\vec r(t)/{\rm d}t$.   
The observations reported here are of the Doppler shifts of known frequencies or pulse rates, 
which are given by
\begin{equation}
f' = f_s \sqrt{\frac{1-(\vec v\cdot \hat e)/c}{1+(\vec v \cdot \hat e)/c}}
\label{EQ1}
\end{equation}
where  $f'$ is the observed frequency, $f_s$ is the frequency at the source, $c$ is the speed of light, 
and $\vec v \cdot \hat e$ is the relative velocity between the source and the observer projected onto 
the direction of observation $\hat e$.   If the source frequency is known, Eq.~\ref{EQ1} can be inverted 
to obtain the relative projected velocity of the object under observation.  

Two methods are used to determine the Earth's velocity.  
The first is based on the observation of 21-cm (\HI) emissions from galactic hydrogen.
A second method is based on precision timing measurements of the pulses from PSR J0332+5434,
which exhibits a pulse rate of $\sim$\,1.4~Hz.    
The pulsar method is surprisingly accurate given the relatively modest nature of our antenna.    By leveraging 
the extreme stability of pulsar rotation rates and the 
exquisite timing accuracy that is publicly available from Network Time Protocol (NTP) servers, 
the necessary timing precision can be obtained with relative ease.    

\section{Apparatus and data acquisition}
The instrument used in this study is a 12.8-m parabolic antenna, located near 
\ifanon 
XXXX, XXXX.
\else 
Carp, Ontario.
\fi  
The antenna is outfitted with a dual-polarization feed system, with suitable low-noise amplifiers 
at the feed-point, as is conventional.    The signals from the feed sub-system are presented to 
a software defined radio (in this case, an Ettus Research N310), and then processed with a multi-CPU Xeon server.  
The reference clock for the receiver is 
locked to a GPS-based precision oscillator, providing a frequency accuracy of better than one part in $10^{8}$.  

Daily transit scan observations are made, with the antenna fixed at an angle 9.2$^\circ$ below the zenith in a 
northerly direction along the north-south meridian.   This direction was chosen to correspond to the daily meridian 
transit of PSR J0332+5434.  Pointing the antenna along the meridian avoids Doppler effects associated with Earth's 
rotational motion.   Since PSR J0332+5434 lies near the galactic plane, a strong \HI peak is also observed.
   
For the pulsar observations, the data take the form of a time series of the total power integrated over 
the observation bandwidth of 20.8~MHz, centered at 1416.4~MHz.    Time samples are spaced by 
1.54~ms.   The small frequency dispersion ($\pm 800~\mu{\rm s}$ over the receiver pass band) 
is neglected. 

\section{Data Analysis}

\begin{figure}[htbp]
   \centerline{\includegraphics[width=0.48\textwidth]{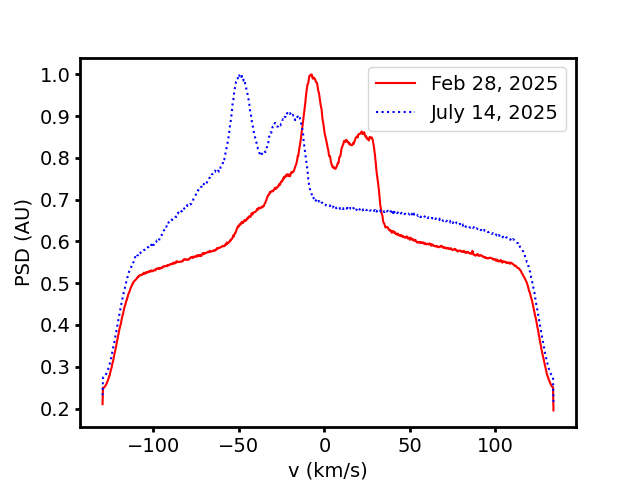}}
   \caption{Raw data Doppler red shift for \HI emissions.  }  
   \label{FIG:H1_vDoppler}
\end{figure}

For the \HI observations,  individual spectra, integrated for a period of five minutes, are taken once per day at the 
time of transit of the target region.   Two such spectra, taken at times separated by  
136 days are shown in Fig.~\ref{FIG:H1_vDoppler}.    The relative shift of the two spectra is a consequence of the change 
in Doppler velocity that results from Earth's orbit.   (The velocity of the gas cloud under observation with respect to 
the solar system barycenter is effectively constant on the timescale of our observations.) 

\begin{figure}[htbp]
   \centerline{\includegraphics[width=0.48\textwidth]{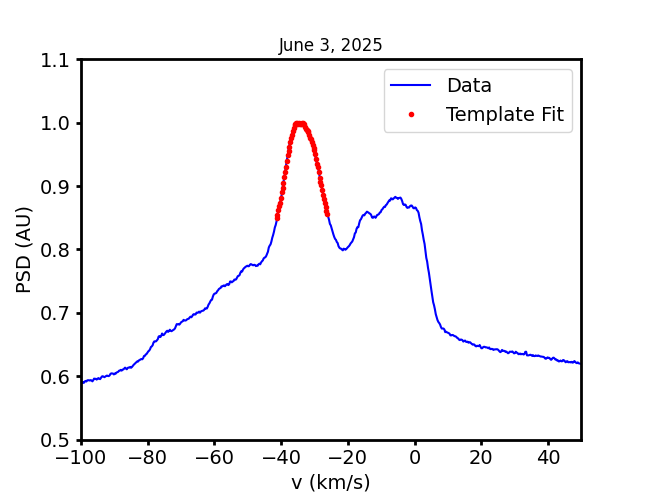}}
   \caption{\HI emission peak shown with fitted peak template. }  
   \label{FIG:H1_fit}
\end{figure}

The Doppler velocity corresponding to the center of the main peak is extracted by fitting a template peak shape to the 
region of the spectrum surrounding the peak.    The center, height, and baseline of the template shape are adjusted to 
provide the best match to the data.    A sample fit is shown in Fig.~\ref{FIG:H1_fit}.   The uncertainty on the 
velocities obtained in this way is estimated from an analysis of adjacent measurements to be 
$\sigma^{\rm \HI}_v \simeq 15.4~{\rm m/s}$.

For the pulsar observations, a folding method is used to make an initial determination of the pulse period and phase 
at the time of observation.    The phase $\phi(t)$ of the pulsar signal consists of a whole number part, which 
corresponds to the number of rotations of the pulsar under observation and a fractional part.     As a general rule, 
only the fractional part of the phase is measured, although the number of turns between the daily observations can 
be inferred if the rotational period is known with sufficient accuracy.      

\begin{figure}[htbp]
   \centerline{\includegraphics[width=0.50\textwidth]{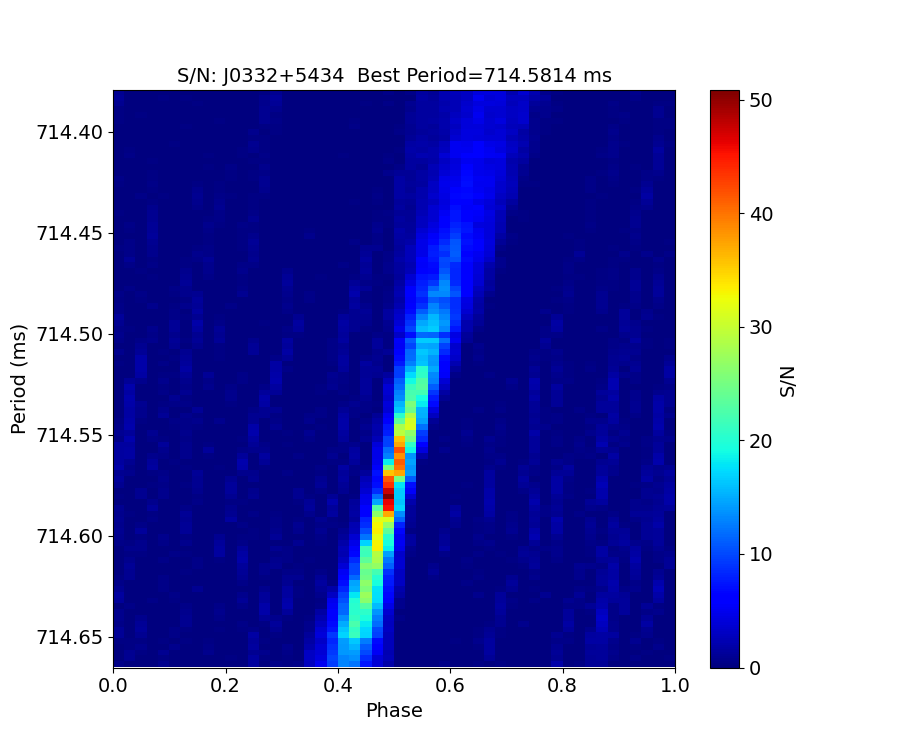}}
   \caption{A sample phase plot from a daily observation.  }  
   \label{FIG:S2NvPeriod}
\end{figure}

The analysis proceeds by dividing the time series into data segments corresponding 
to an assumed period and then averaging the stacked segments in such way that the pulsar signal 
emerges from the noise when the assumed period corresponds to the true period of the signal at the receiver input. The 
assumed pulse period, $T$,  is varied in 100 steps over the range 
$T_0(1 - \varepsilon) < T < T_0(1 + \varepsilon)$, where $T_0$ is the pulse period of the 
pulsar in a frame corresponding to the barycenter of the solar system, and the 
parameter $\varepsilon = 2 \times 10^{-4}$.     Fig.~\ref{FIG:S2NvPeriod} shows the signal-to-noise ratio 
as a function of the fractional part of the pulse phase and the trial period for a selected daily observation.

To determine the average pulse frequencies between a pair of daily measurements, we use
\begin{equation}
\phi_{i+1} =  \phi_i + \int_{t_i}^{t_{i+1}} f(t) dt  
 = \phi_i + \left(t_{i+i} - t_{i} \right) \bar f ,
\label{EQ:freq1}
\end{equation}  
where $\phi_i$ and $t_i$ are the phase and time of the $i^{\rm th}$ measurement, and $\bar f$, which is the quantity 
we seek to measure, is the average pulsar frequency between the measurements.   Since our roughly ten-minute observations 
are spaced by one sidereal day (23.93~hrs) we cannot directly measure $\bar f$.    We can, however, infer the value of $\bar f$ 
with an iterative procedure.   Specifically, Eq.~\ref{EQ:freq1} can be recast  as 
\begin{equation}
\phi_{i+1} - \phi_i 
= \left({\bar f}_0 + {\Delta \bar f}\right)\left(t_{i+1} - t_i\right),
\label{EQ:freq2}
\end{equation}
where ${\bar f}_0$ is an initial estimate for $\bar f$, and $\Delta \bar f$ is a correction to that estimate.  Since the 
fractional parts of $\phi_{i+1}$ and $\phi_i$ and the times $t_{i+1}$ and $t_i$  
are measured,  if we have a good estimate for the value ${\bar f}_0$, we can use Eq.~\ref{EQ:freq2} to find $\Delta \bar f$, 
from which $\bar f$ can be determined using 
$\bar f = \bar f_0 + \Delta \bar f$.     Phase measurement errors on the order 
of $\delta \phi \sim 0.0011$ are typically obtained, resulting in a velocity uncertainty given by 
\begin{equation}
 \delta v = \sqrt{2}\, \frac{\delta \bar f}{f} c  =
  \sqrt{2}\, \frac{\delta \phi }{\Delta t} \frac{c}{f}  \sim 4~{\rm m/s},
 \end{equation}
where $\Delta t = t_{i+1} - t_i$ is one sidereal day, $c$ is the speed of light, and $f\simeq1.4$~Hz is the pulse rate.  
(Observed values of $\delta \phi$, which depend on the signal-to-noise ratio, vary considerably from day to day, owing 
to the variation in pulsar signal strength that results from scintillation effects in the interstellar 
medium\cite{Stinebring96,Brown25}.) 

A key to this method is a sufficiently accurate initial value for ${\bar f}_0$.   If ${\bar f}_0$ differs from the true 
value $\bar f$ by too much, a phase ambiguity can result (recall that we measure only the fractional part of the $\phi_i's$).    
To avoid the ambiguity, we require 
\begin{equation}
\left(\bar f - \bar f_0 \right)\left(t_{i+1} - t_i\right) < 0.5~.
\end{equation}
Numerically, this means that $\delta f$, the error in the initial estimate for the frequency,  must satisfy
\begin{equation}
\delta f = \left| \bar f - \bar f_0 \right| < \frac{1}{2\left(t_{i+1} - t_i\right)} \sim 6~\mu{\rm Hz}, 
\label{EQ6}
\end{equation}
where we have taken the time between measurements to be one sidereal day $\simeq 8.6 \times 10^4$~s.    
Although this is somewhat smaller than the $\sim 20~\mu{\rm Hz}$ measurement uncertainty that can be reliably 
obtained in a single 10-minute observation (see Fig.~\ref{FIG:S2NvPeriod}), there are other methods that can 
be used to achieve the requisite accuracy.   These will be discussed in Section~\ref{SEC:pulsar_fits}.

\section{Orbital Model}
Both the \HI and pulsar data are fit to an orbital model comprising multiple parameters.   In this model, 
the Earth-Sun distance is given by      
\begin{equation}
r(\theta) = \frac{a\left(1 - e^2\right)}{1 + e \cos\left(\theta - \theta_0\right)},
\label{EQ3}
\end{equation}
where: $a$ is the semi-major axis of Earth's orbit, $e$ is the orbital eccentricity, and $\theta_0$ is the 
perihelion angle of the orbit.   A fourth parameter $t_0$ takes into account the phase of the orbit.  For orbits 
with $e\ne0$, $t_0$ is taken to be the time at which Earth passes through perihelion.   For the \HI data, a fifth 
parameter $v_{\rm offset}$ is used to take into account the overall Doppler offset of the \HI peak under observation.  
For some of the pulsar fits, a velocity-offset parameter, which takes into account possible small inaccuracies in the $f_s$ values, is
included.  

\begin{figure}[htbp]
   \centerline{\includegraphics[width=0.35\textwidth]{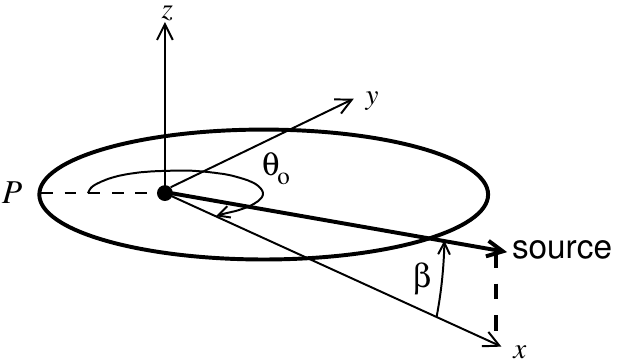}}
   \caption{Schematic view of Earth's orbit.   The orbit, which is drawn with exaggerated eccentricity,  
   lies in the $x$-$y$ plane.      }  
   \label{FIG3}
\end{figure}

We choose a coordinate system in which the orbit lies in the $x$-$y$ plane, and the star (or gas cloud) under observation 
lies at an angle $\beta$ above the $x$-axis, as shown in Fig.~\ref{FIG3}.   The time development of $\theta(t)$ is 
determined using an iterative solution of Kepler's equation $M(t) = \theta(t) - e\sin\theta(t)$, where $M(t)=2\pi/T_{\rm year}$ 
is the mean anomaly and $\theta(t)$ is the eccentric anomaly (Earth's position).    

In principle, $\beta$, which is referred to as the ecliptic latitude, could be included as a parameter in the fit.   In practice, 
however, the observed Doppler velocity is given by $v^{\rm obs}_x = v_x \cos\beta$  and $v_x$ scales linearly with $a$, meaning 
that $a$ and $\cos\beta$ are strongly correlated.   As a result, it is not possible to simultaneously fit for both $a$ 
and $\beta$.   We therefore keep $\beta$ fixed in the fits.   For the \HI observations, we use a value of 
$\beta_{\rm \HI}=(33.668\pm 0.095)^\circ$, corresponding 
to the direction in which the telescope was pointing at the time of the observations 
 (${\rm RA} = 55.495 ^\circ$ and ${\rm dec}= 54.501^\circ$).    The error on $\beta_{\rm \HI}$ is estimated by varying the RA 
by an amount corresponding to one fourth of the total change in RA during the transit. For the pulsar data, we use a fixed value of 
$\beta_J=(34.2629 \pm 0.0018)^\circ $ taken from other observations\cite{Manchester05,Hinton06}.   The error 
on $\beta_J$, is conservatively estimated to be one half of the change due to the proper motion of J0332+5434 since the year 2000.
\section{Results}
\subsection{\HI Fits}
\begin{figure}[htbp]
   \centerline{\includegraphics[width=0.48\textwidth]{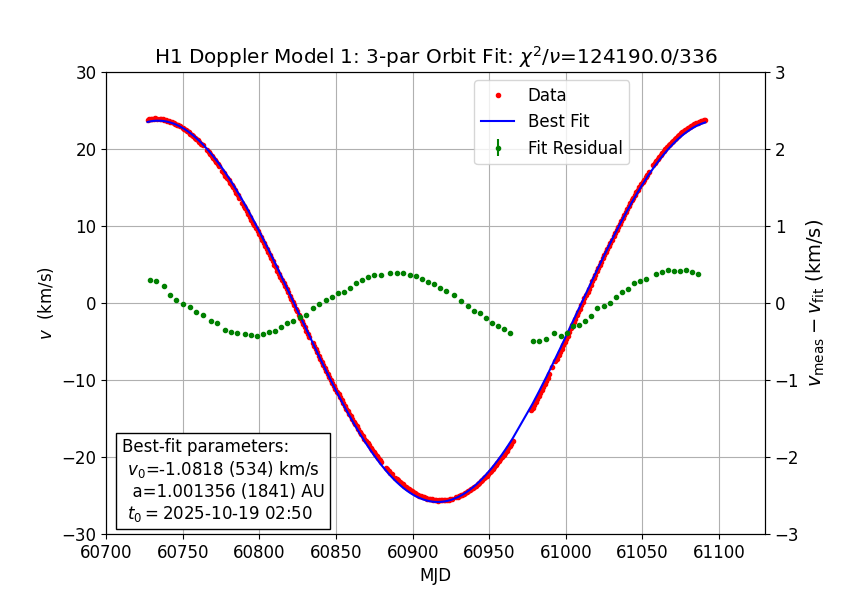}}
   \caption{Fit of \HI data to circular orbit for one year of data taking.     
   The velocities are plotted as a function of Modified Julian Day (MJD).  The blue points represent the 
   data and the red curve is the best fit and use the vertical scale on the left.    The green points are 
   the residual difference between the data and the fit and use the vertical scale on the right.   The error 
   bars are smaller than the points.   The numbers in parentheses following the best-fit parameter values 
   correspond to the last digits of the statistical error from the fit. }  
   \label{FIG4}
\end{figure}

An initial fit of the \HI data to a model that assumes a circular orbit ($e=0$) is shown in Fig.~\ref{FIG4}.  A circular model provides 
qualitative agreement with the data and yields a semi-major axis value of $a = 1.0014 \pm 0.0018$, which is consistent with 
the established value, but as a quantitative matter, the fit is poor.    This can be seen in the residuals plot, where a large 
sinusoidal variation having a period of 6 months is observed.   
\begin{figure}[htbp]
   \centerline{\includegraphics[width=0.48\textwidth]{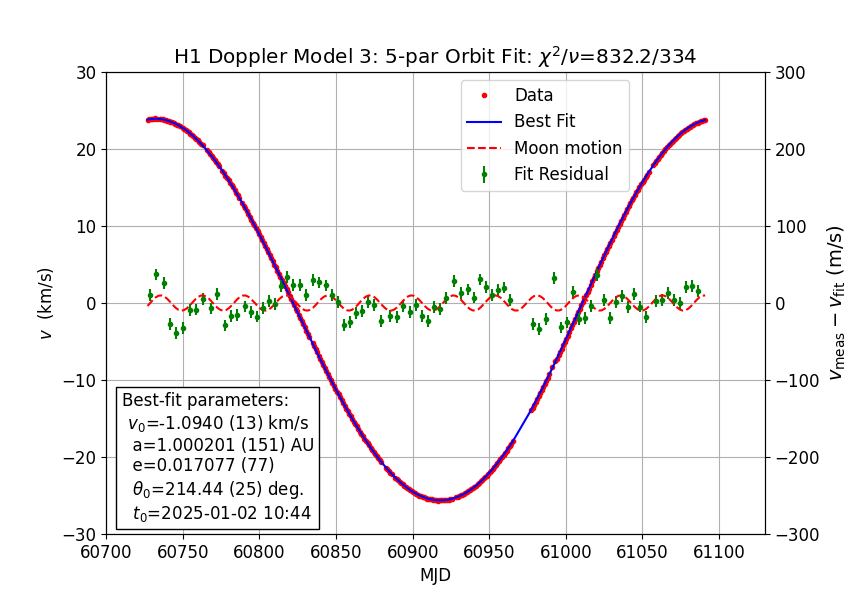}}
   \caption{Fit of \HI data to an elliptical orbit for one year of data taking.    The blue points 
   represent the data and the red curve is the best fit (left-hand scale).   The residuals are plotted in green (right-hand scale).  }  
   \label{FIG5}
\end{figure}
Figure~\ref{FIG5} shows a fit when eccentricity is added to the model.   In addition to the eccentricity $e$, a parameter $\theta_0$, 
which corresponds to the orientation of the major axis of the ellipse, is included in the fit.    
The quality of the fit is greatly improved, although the value of the $\chi^2$ per 
degree of freedom, which is $\chi^2/\nu = 2.49$, suggests either that the data points are plagued by systematic errors 
or that there is a problem with the model.    One possible problem is a shift in the apparent position of the \HI peak that results 
from gain variations across the receiver pass band.   In addition, the fit model does not include the motion of the Earth 
associated with the Moon, which is approximately $11$~m/s.   Including the effect of the Moon motion lowers the reduced $\chi^2$ 
modestly to $\chi^2/\nu = 2.44$. 

Two main sources of systematic error are considered.    One is the aforementioned shift in the peak position resulting from 
pass band gain variations.   This  is estimated by simulating the effect of a parabolic ``bowing'' of the gain across the 
receiver pass band.  If the amount of bowing results in a gain that is $\pm 2.5\%$ higher at the center of the pass band 
relative to the edges,  a $\pm 0.0017$ impact on $a$ and a $\pm 0.000018$ impact on $e$ result.   A second source of systematic 
error is the value of the ecliptic latitude $\beta_{\rm \HI}$, which is uncertain because the precise coordinates of the \HI 
cloud under observation are not known.  We take the uncertainty in $\beta_{\rm \HI}$ to be $0.05^\circ$, which is one quarter 
of the change in $\beta_{\rm \HI}$ from the beginning to the end of the drift scan.   The corresponding uncertainty in 
$a$ is $\pm 0.0012$, and the impact on $e$ is negligible.   Combining these systematic errors in quadrature yields
\begin{equation}
\sigma^{\rm sys}_a = \pm 0.0021 \quad {\rm and} \quad \sigma^{\rm sys}_e = \pm 0.000018 .
\end{equation} 

\subsection{Pulsar Fits}
\label{SEC:pulsar_fits}
Velocity values are computed from the observed frequencies using Eq.~\ref{EQ1} with $f_s$ values---i.e., values referred to the barycenter of the solar system---taken from the literature\cite{Hobbs04,Cassity25}.    As a result of pulsar spindown, the $f_s$ values decrease by $\Delta f_s = 147$~nHz (equivalent to $\Delta v=32$~m/s) over the 451-day observing interval. 

\begin{figure}[htbp]
   \centerline{\includegraphics[width=0.48\textwidth]{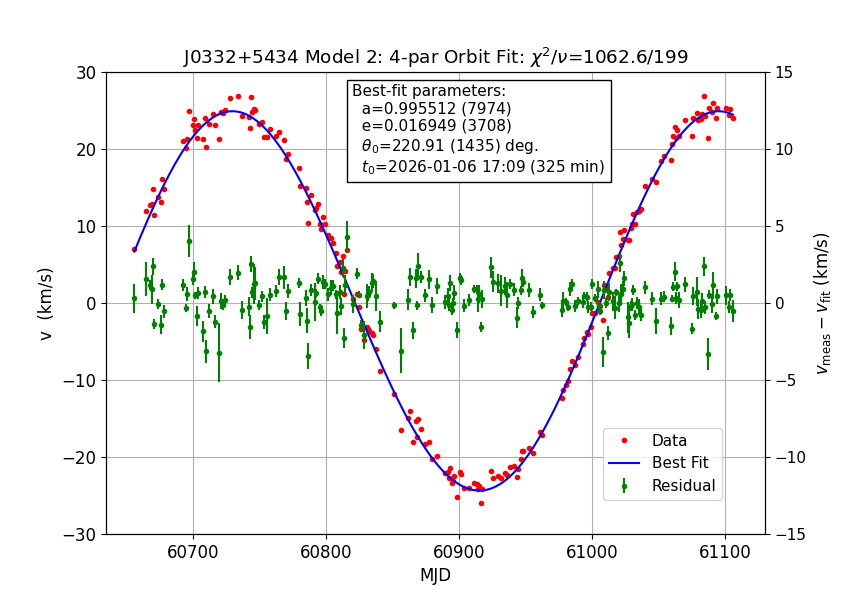}}
    \caption{Fit of pulsar data to an elliptical orbit.   The data points are based on individual daily measurements, 
    having an average velocity uncertainty of  540~m/s.  The blue points represent the data and the red curve is the 
    best fit (left-hand scale).   The residuals are plotted in green (right-hand scale).  }  
   \label{FIG6}
\end{figure}

\begin{figure}[htbp]
   \centerline{\includegraphics[width=0.48\textwidth]{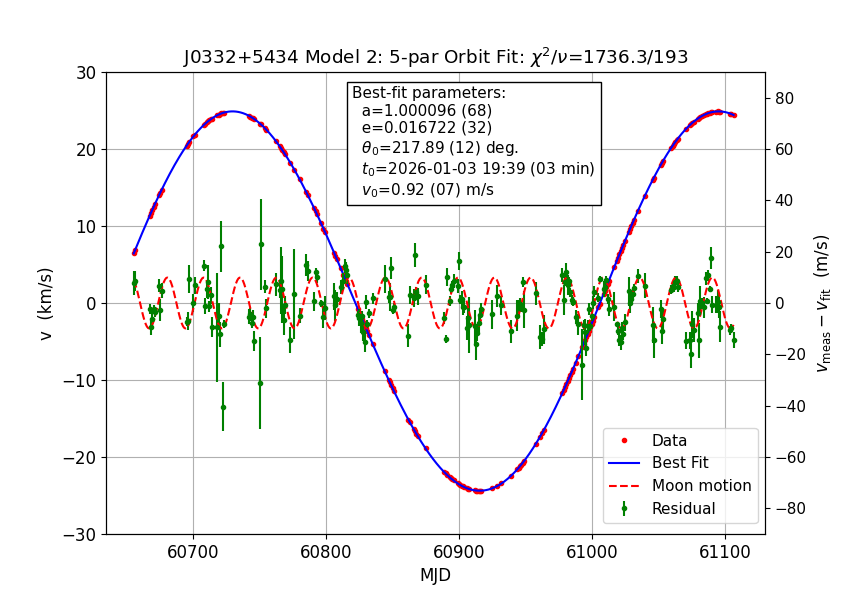}}
 \caption{Fit of pulsar data to an elliptical orbit without taking into account the effects of the Moon.   The data points are based on pairs of measurements separated by at least one day and having an average velocity uncertainty of  4.03~m/s.   The blue points represent the data and the red curve is the best fit (left-hand scale).   The residuals are plotted in green (right-hand scale).  }  
   \label{FIG7}
\end{figure}

\begin{figure}[htbp]
   \centerline{\includegraphics[width=0.48\textwidth]{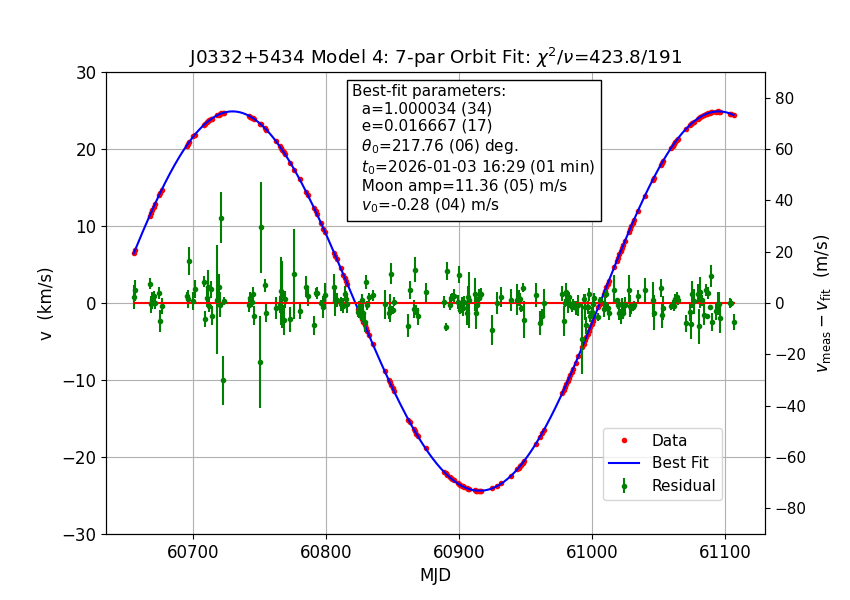}}
 \caption{Fit of pulsar data to an elliptical orbit with the effects of the Moon taken into account.  The data points are based 
 on pairs of daily measurements.   The blue points represent the data and the red curve is the best fit (left-hand scale).   
 The residuals are plotted in green (right-hand scale).  }  
   \label{FIG8}
\end{figure}

As noted above, extraction of precise frequency values from a pair of daily measurements using Eq.~\ref{EQ:freq2} requires 
a sufficiently precise initial estimate for $\bar f_0$.   Although the accuracy of a ten-minute single-day observation falls 
short of what is needed, the average of several such measurements is sufficient.    A suitable average 
can be obtained by fitting all of the single-day measurements to the orbital model.   The results of such a fit are shown 
in Fig.~\ref{FIG6}.   The fitted parameters yield $\bar f_0$ values that are accurate to 
within $\pm 0.6~\mu$Hz, which is much smaller than the accuracy set forth in Eq.~\ref{EQ6}.
The frequency values obtained using Eq.~\ref{EQ:freq2} based on the $\bar f_0$ initial estimates from the fit shown in 
Fig.~\ref{FIG6} are used to generate a new set of highly accurate frequencies, which are re-fitted to the orbital model.   
Figure~\ref{FIG7} shows a fit that uses the more precise frequency values, but does not take into account the gravitational 
pull of the Moon on the Earth,  the effects of which are evident in the residuals plot.    In Fig.~\ref{FIG8}, two fit 
parameters are added to account the motion of the Earth that results from the Moon.   The quality of the fit improves considerably.   

The $\chi^2$ per degrees of freedom for the fit of Fig.~\ref{FIG8} is 424/191.    The corresponding $p$-value is of 
order $10^{-16}$, which clearly cannot be accounted for as a statistical fluctuation: either the fit model is wrong 
or the errors on the data points have been underestimated.    To test the former hypothesis,  simulated data extracted 
from JPL's Horizon API\cite{Giorgini96} were fitted to the model.   The resulting fit residuals exhibited the same period 
as the Moon's orbit and had a residual RMS of $\sigma_{\rm model} =0.86$~m/s.   This result is not surprising, since the 
fit model assumes that the Moon's orbit is circular.    To take this mis-modeling into account,  $\sigma_{\rm model}$ was 
added in quadrature to the errors on the data points.    Doing so yields $\chi^2/\nu = 368/191$, which though improved, 
is still too large to attribute to statistics.     Inspection of the contribution of individual points to the $\chi^2$ 
suggests that some or all of the problem may be attributable to an underestimate of the errors for the points with 
the smallest errors.     Adding 3.0~m/s in quadrature to the data-point errors produces a $\chi^2/\nu=190/191$, with a 
corresponding $p$-value of $\sim0.52$, and has only a small impact on the fitted values.

Table~\ref{TAB1} shows a comparison between the parameters extracted from the fit shown in Fig.~\ref{FIG8} and the 
established (accepted) values.   The impacts on $a$ and $e$ for four main sources of systematic error are summarized 
in Table~\ref{TAB2}.  The $\beta_J$ entry reflects the effect of changing the pulsar's ecliptic latitude by one standard 
deviation.   The  ``Clock model'' uncertainty reflects the difference in fitted values when changing from the pulsar clock 
parameterization of Ref.~\cite{Cassity25} (the one used in the analysis) to the older parameterization of  Ref.~\cite{Hobbs04}.
The Moon orbit term is taken to be the change in the fitted 
parameters resulting from adding a term corresponding to the Moon motion to the observational data.  
The Jupiter orbit term is estimated by correcting the data by an amount corresponding to the calculated effect of Jupiter 
on Earth's orbit around the Sun.  

   The best-fit values of $a$ and $e$ are consistent with the accepted values within their statistical errors.   The values 
for the other parameters are in qualitative agreement, although there are differences that significantly exceed the best-fit 
statistical errors.   The agreement on the time of perihelion $t_0$ is likely fortuitous; the fit is made with the 
assumption that the orbit is closed with a period of 365.2422 days.     This is, however, only approximately true, 
since the period of the lunar orbit is not commensurate with the period of Earth's orbit around the Sun, and the day 
of perihelion varies between January 2 and January 5.   The gravitational effects of other planets also play a role.   

\begin{table}[htbp]
\centering
    \caption{Comparison between best-fit values from Fig.~\ref{FIG8} and the established values.   The stated errors 
    are statistical only.}
    \label{TAB1}
    \begin{tabular}{ccc}
        \toprule
        \textbf{Parameter} & \textbf{Best Fit} & \textbf{Accepted} \\
	 \hline 
        $a$    & $1.000034 \pm 0.000034$     & 1.000000     \\
        $e$    & $0.016667 \pm 0.000017$     & 0.016698   \\
       $\theta_0$   & $(217.76 \pm 0.06)^\circ$     & 217.09$^\circ$    \\
       $t_0$   & 2026-01-03 16:29  &   2026-01-03 16:59 \\
         \hline 
    \end{tabular}
\end{table}

\begin{table}[htbp]
\centering
    \caption{Effect of systematic uncertainties on the fitted values of $a$ and $e$ for the pulsar fits.}
    \label{TAB2}
    \begin{tabular}{ccc}
        \toprule
        \textbf{Source} & $a$ & $e$ \\
	 \hline 
        $\beta_J$    &  0.000038 &  $\sim 0$ \\
        Clock model   &     0.000048  & 0.000003 \\
       Moon  &   0.000021 & 0.000011    \\
       Jupiter & 0.000022 & 0.000008 \\
       \hline 
      Total   & 0.000068 & 0.000014  \\
         \hline 
    \end{tabular}
\end{table}

\section{Discussion}
Although the \HI measurements presented here benefit from the relatively good angular resolution of the 12.8-m dish,  
the accuracy obtained from a smaller dish (2$\sim$3-m-diameter) should be more than sufficient for pedagogical purposes.    
Many university teaching programs have access to dishes of this size. 

The pulsar measurements can be viewed in the context of the current-day NANOGrav\cite{Agazie23} project, 
which seeks evidence of low-frequency gravitational waves and other interesting phenomena through precision 
timing of pulsars.   The results presented here can be viewed as a reverse engineering of the codes used by NANOGrav 
to take into account timing shifts resulting from the orbital motion of the Earth\cite{Susobhanan24}.   

Since access to a 12.8-m or larger dish lies beyond the reach of most teaching programs, we have provided 
the raw frequency measurements in the supplementary material, the analysis of which can serve as the basis 
for an intermediate level laboratory exercise.

\begin{acknowledgments}
We gratefully acknowledge the contributions of 
\ifanon
XXXXX
\else 
Allan Duncan of CSS Building Inc., Justin Mutter of the 
Carp Observatory, as well as Bill Wagstaff and Ken Whitnall.    We thank Curtis Macchioni for useful 
comments on the manuscript. 
\fi 
\end{acknowledgments}

\end{document}